\documentclass[preprint,showpacs,preprintnumbers,amsmath,amssymb]{revtex4}
\usepackage{graphicx}
\begin{document}
\title{Theory of Phonon-Drag Thermopower of Extrinsic Semiconducting Single-Wall Carbon
Nanotubes and Comparison with Previous Experimental Data}

\author{M. Tsaousidou}
\email{rtsaous@upatras.gr}
\affiliation{Materials Science
Department, University of Patras, Patras 26 504, Greece. }

\date{\today}

\begin{abstract}
A theoretical model for the calculation of the phonon-drag
thermopower, $S^{g}$, in degenerately doped semiconducting
single-wall carbon nanotubes (SWCNTs) is proposed. Detailed
calculations of $S^{g}$ are performed as a function of
temperature, tube radius and position of the Fermi level. We
derive a simple analytical expression for $S^{g}$ that can be
utilized to determine the free carrier density in doped nanotubes.
At low temperatures $S^{g}$ shows an activated behavior
characteristic of the one-dimensional (1D) character of carriers.
Screening effects are taken into account and it is found that they
dramatically reduce the magnitude of $S^{g}$. Our results are
compared with previous published experimental data in bulk p-doped
SWCNT materials. Excellent agreement is obtained in the
temperature range 10-200~K for a consistent set of parameters.
This is a striking result in view of the complexity of these
systems.

\end{abstract}

 \maketitle

\section{Introduction}
Thermopower, $S$, is an important transport coefficient that
offers valuable information about the electronic structure, the
scattering processes and the mechanisms of carrier-phonon coupling
in a system. In the last few years there has been growing
experimental interest in $S$ of single wall carbon nanotubes
(SWCNTs). Several groups have reported thermopower measurements on
bulk SWCNT materials (e.g., mats, fibers, films)
\cite{Grigorian,Hone2,Hone,Collins,Bradley,
Mahan2,Sumanasekera,Fischer03,Fischer05} and on individual
SWCNTs~\cite{Small,Llaguno,Yu}. However, only modest progress has
been made up to now in understanding the unique features of $S$ in
these systems. Interesting issues concerning the large positive
thermopower ($\sim$ 80~$\mu$V/K) in pristine samples
\cite{Hone2,Collins,Bradley,Sumanasekera,Fischer03}, the change of
sign of $S$ upon exposure to oxygen~\cite{Collins,Sumanasekera}
and the effect of carrier-phonon coupling
\cite{Mahan2,Fischer03,Fischer05,Mahan,Mahan13,Mahan23} on $S$
still remain open.

$S$ consists of two additive contributions which are diffusion,
$S^{d}$, and phonon-drag, $S^{g}$. $S^{d}$ is due to the carrier
diffusion in the presence of a temperature gradient and for
degenerate systems varies linearly with $T$ according to Mott's
expression. $S^{g}$ originates from the interchange of momentum
between acoustic phonons and carriers via the carrier-phonon
interaction. The first theoretical models for the study of the
phonon drag in metals~\cite{Bailyn} and
semiconductors~\cite{Herring} were developed half a century ago.
More recently, extensive theoretical and experimental work has
been carried out on $S^{g}$ of low-dimensional semiconductor
structures~\cite{Butcherbook,Fletcher04,Tsaousidoubook}.

Recent experiments on $S$ in p-doped SWCNT films and
fibers~\cite{Fischer03,Fischer05} provided clear evidence for the
presence of $S^{g}$ at $T>15-20$~K. On the theory level, however,
there is still an ongoing discussion about the role of $S^{g}$ in
measured thermopower~\cite{Mahan13,Mahan23}. So far, the
theoretical studies of $S^{g}$ are confined to metallic armchair
(10,10) tubes~\cite{Mahan,Fischer03}. However, in perfect metallic
tubes with mirror electron-hole symmetry both
$S^{d}$~\cite{Mahan2} and $S^{g}$~\cite{Mahan13,Mahan23} are
expected to be negligibly small compared to the experimental data,
due to the competition between the opposite contributions of
electrons and holes. We note that the accuracy of the existing
theoretical models~\cite{Mahan,Fischer03} for $S^{g}$ in metallic
tubes has been questioned recently by Mahan~\cite{Mahan23}. Also,
a recent theoretical work~\cite{Kuroda} pointed out that
thermopower vanishes in one-dimensional conductors with a linear
energy dispersion (as in the case of metallic tubes) due to
electron-hole symmetry.

In this paper we propose a theoretical model for the phonon-drag
thermopower in semiconducting SWCNTs that are characterized by a
non-linear energy dispersion. (A brief discussion on the behavior
of $S^{g}$ in this kind of nanotubes appears in
[\onlinecite{Tsaousidoubook}].) We suggest that the measured
thermopower in doped samples is due to the contribution of
degenerate semiconducting nanotubes. In our model $S^{g}$
originates from carrier-phonon intraband scattering within the
first 1D subband. As we discuss below, the dominant contribution
to $S^{g}$ is made by long-wavelength acoustic phonons that
backscatter carriers across the Fermi surface. In this case the
carrier-phonon coupling is much weaker in metallic tubes than in
semiconducting tubes~\cite{Ando} and, consequently, $S^{g}$ is
expected to be substantially larger in the latter ones.

We note that upon chemical or electrostatic doping the Fermi level
can be pushed into the conduction or valence band and the
degenerate semiconducting tubes can be considered as
one-dimensional metals. Therefore the terms \lq\lq metallic\rq\rq
and \lq\lq semiconducting\rq\rq refer only to the different
electronic structure in the two types of tubes (see, for example,
Ref.[\onlinecite{Charlier})].

There are two equivalent theoretical approaches to the problem of
phonon drag\cite{Tsaousidoubook}. In the first approach phonons
are perturbed in the presence of a weak temperature gradient
$\nabla T$. Non-equilibrium phonons transfer part of their
momentum to carriers due to the carrier-phonon coupling. Then the
phonon-drag contribution to the thermoelectric current
$J^{g}=L^{g}\nabla T$ is calculated by solving the coupled
Boltzmann equations for carriers and acoustic phonons
\cite{Mahan,Bailyn,CB,Kubakaddi}. The phonon-drag thermopower is
readily obtained by $S^{g}=-L^{g}/\sigma$ where $\sigma$ is the
carrier conductivity. In the second approach carriers are
accelerated isothermally in the presence of a weak electric field
${\bf E}$ and impart some of their momentum to phonons due to the
carrier-phonon coupling. Then the resulting phonon heat current
and the phonon-drag contribution to the Peltier coefficient is
calculated
\cite{Herring,Puri,Gerin,Kubakaddi89,Lyo89,Fromhold,Tsaousidou2}.
This method of evaluating $S^{g}$ is referred as
$\Pi$-approach~\cite{Herring} because it provides a direct
estimation of the Peltier coefficient. The equivalence of the
above two approaches is secured by Onsager's symmetry relation. In
this paper we follow the second approach which is more general and
it can be applied even in systems where carriers do not behave
semiclassically~\cite{Kubakaddi89,Lyo89,Fromhold,Tsaousidou2}.

The paper is organized as follows. In Sec.II we introduce the
theoretical model for the calculation of $S^{g}$ in the
semiclassical transport regime. An explicit expression for $S^{g}$
is derived in Sec.IIB and in Sec.IIC we derive a simple
approximate expression for $S^{g}$ for the case of a highly
degenerate semiconducting tube. Numerical results for $S^{g}$ as a
function of temperature, tube radius and position of Fermi level
are presented in Sec.III. In the same Section we discuss the
effect of screening. In Sec.IV we compare our theory with
available experimental data for acid-doped bulk SWCNT samples.

\section{Theory}
\subsection{Description of the physical system}
We assume that the nanotube is a long indefinitely thin cylinder
of radius $R$ and length $L$. The nanotube axis is along the
$z-$direction. The carrier wave function is~\cite{Lin}
\begin{equation}
\label{wavefunction} \Psi_{l\,k}({\bf
r})=\frac{1}{\sqrt{L}}e^{i\,k\,z}\frac{1}
{\sqrt{2\pi}}e^{i\,l\,\theta} \frac{1}{\sqrt{R}}\delta(r-R),
\end{equation}
where, ${\bf r}$ is the space vector, $k$ is the carrier wave
vector along the axial direction, $\theta$ is the azimuthal angle
and $l$ labels 1D orbital subbands associated with the carrier
confinement along the circumference. We assume, that the Fermi
level, $E_{F}$, is located between the first and the second 1D
subbands (i.e., only the ground subband is occupied). Then, the
carrier energy is
\begin{equation}
E_{k}=E_{1}+\frac{\hbar^{2}k^{2}}{2m^{*}}
\end{equation}
where $m^{*}$ is the carrier effective mass and $E_{1}$ denotes
the position of the first van Hove singularity.

In carbon nanotubes phonons also exhibit 1D character. The lattice
displacement at a point ${\bf r}$ is \cite{Ando}
\begin{equation}
\label{latdis} {\bf u(r)}=\hat{{\bf
\eta}}_{mq}e^{iqz}e^{i\,m\,\theta}
\end{equation}
where, $\hat{{\bf \eta}}_{mq}$ is the polarization vector, $q$ is
the phonon wave vector in the axial direction and $m=0,\pm 1, \pm
2, ...$ denotes the phonon modes associated with phonon
confinement along the circumference. Due to the conservation of
angular momentum only the three low-energy acoustic modes with
$m=0$ (the so-called twisting, stretching and breathing modes)
contribute to the carrier-phonon intraband scattering. The phonon
frequencies and polarization vectors have been calculated within
the continuum model proposed by Suzuura and Ando \cite{Ando}.

The carrier-phonon interaction in carbon nanotubes has been
studied in several texts within the tight-binding
approximation~\cite{Jishi,Mahan5,Mahan4,Jiang,Popov} or a
continuous elastic theory~\cite{Ando,Martino2,Pennington2,Ragab}.
Here we follow the continuous model of Suzuura and Ando
\cite{Ando} according to which the carrier-phonon coupling is
described via the acoustic deformation potential
\begin{equation}
\label{def} U({\bf r})=D\left(\frac{1}{R}\frac{\partial
u_{\theta}}{\partial \theta}+\frac{\partial u_{z}}{\partial
z}+\frac{u_{r}}{R}\right),
\end{equation}
where $D$ is the deformation potential constant. The deformation
potential approximation provides a good description of the carrier
interaction with long-wavelength acoustic phonons.  The last term
in Eq.~(\ref{def}) accounts for the nonzero curvature of the
nanotube \cite{Ando}. The twisting mode does not participate to
carrier-phonon scattering via the deformation potential coupling.
Moreover, in the long-wavelength limit ($qR\ll 1$), which is the
regime of our interest, the breathing mode is dispersionless and
does not contribute to $S^{g}$. Thus, in what follows we consider
only the stretching mode which is characterized by a linear
dispersion $\omega_{q}=v_{s}|q|$ where, $v_{s}$ is the sound
velocity. The phonon polarization vector,
$\hat{{\bf\eta}}=(\eta_{\theta},\eta_{z},\eta_{r})$, for this mode
in the limit $qR\ll 1$ is
\begin{equation}
\label{eta} \hat{{\bf \eta}}_{q}=(0,\frac{1}{a},\frac{-i\nu
qR}{a})
\end{equation}
where $a=\sqrt{1+\nu^{2}q^2R^2}$ and $\nu$ is Poisson's ratio.
Ignoring the terms proportional to $q^2R^2$ the above expression
becomes identical with the one derived by De Martino {\em et al}.
\cite{Martino}.

\subsection{An explicit expression for the phonon-drag thermopower}
We assume a small electric field $E$ in the axial direction of the
nanotube. The presence of $E$ creates a net flux of carriers along
the axis of the tube which results in a momentum transfer to
phonons through the carrier-phonon coupling. We calculate the
resulting phonon heat flux $Q$ and obtain the phonon-drag
contribution to the transport coefficient
\begin{equation}
\label{M} M^{g}=Q/E.
\end{equation}
To get $S^{g}$ we utilize the Onsager's relation
\begin{equation}
\label{onsager}
S^{g}=\frac{M^{g}}{T\sigma}
\end{equation}
where $\sigma$ is the carrier conductivity and $T$ the absolute
temperature.

The phonon heat flux is given by
\begin{equation}
\label{Q} Q=\frac{1}{L}\sum_{q}\hbar\omega_{q}v_{q}N^{1}_{q},
\end{equation}
where $v_{q}=v_{s}q/|q|$ is the phonon group velocity and
$N^{1}_{q}=N_{q}-N^{0}_{q}$ is the first order perturbation of the
phonon distribution function.

The perturbation $N^{1}_{q}$ is determined by the steady-state
Boltzmann equation for phonons in the relaxation time
approximation when $\nabla T=0$. Namely,
\begin{equation}
\label{boltzmann}
-\frac{N^{1}_{q}}{\tau_{ph}}+\left(\frac{\partial N_{q}}{\partial
t}\right)_{ph-c}=0,
\end{equation}
where $\tau_{ph}$ is the phonon relaxation time associated with
phonon-phonon collisions and phonon scattering by imperfections.
For simplicity we have ignored the dependence of $\tau_{ph}$ on
$q$. $(\partial N_{q}/\partial t)_{ph-c}$ is the rate of change of
the phonon distribution function $N_{q}$ due to phonon scattering
by carriers. It is written in the standard form
\begin{eqnarray}
\label{ph-c} \left(\frac{\partial N_{q}}{\partial
t}\right)_{ph-c}=g_{s}g_{v}\sum_{k,k^{\prime}}f_{k^{\prime}}
(1-f_{k})P_{q}^{e}(k^{\prime},k)
\nonumber\\-f_{k}(1-f_{k^{\prime}})P_{q}^{a}(k,k^{\prime}),
\end{eqnarray}
where $g_{s}$ and $g_{v}$ are the spin and the valley
degeneracies, respectively, $f_{k}$ is the carrier distribution
function and $P_{q}^{a(e)}(k,k^{\prime})$ are the transition rates
at which the carrier in a state $k$ is promoted to a state
$k^{\prime}$ by absorbing (emitting) one phonon with wave vector
$q$.

When the external field $E$ is weak Eq.~(\ref{ph-c}) is linearized
and is solved in terms of $N^{1}_{q}$. Then we get
\begin{eqnarray}
\label{N1nn} \left(\frac{\partial N_{q}}{\partial
t}\right)_{ph-c}&=&-\frac{N^{1}_{q}}{\tau_{pc}(q)}\nonumber\\
&+&\frac{g_{s}g_{v}}{k_{B}T}\sum_{k,k^{\prime}}
\Gamma_{k^{\prime},k}\left(\frac{f^{1}_{k}}{df^{0}_{k}/dE_{k}}-
\frac{f^{1}_{k^{\prime}}}{df^{0}_{k^{\prime}}/dE_{k^{\prime}}}\right),\nonumber\\
\end{eqnarray}
where, $\tau_{pc}(q)$ is the phonon relaxation time associated
with scattering by carriers given by
\begin{equation}
\tau_{pc}^{-1}(q)=g_{s}g_{v}\sum_{k,k^{\prime}}
\Gamma_{k^{\prime},k}/[N^{0}_{q}(N^{0}_{q}+1)],
\end{equation}
and $\Gamma_{k^{\prime},k}$ is the average equilibrium rate of
absorption of phonons with wave vector $q$. It is given by
\begin{equation}
\label{Gamma}
\Gamma_{k^{\prime},k}=f^{0}_{k}(1-f^{0}_{k^{\prime}})P^{a0}_{q}(k,k^{\prime}),
\end{equation}
where $f_{k}^{0}\equiv
f^{0}(E_{k})=\{\exp[\beta(E_{k}-E_{F})]+1\}^{-1}$ (with
$\beta=1/k_{B}T$) is the Fermi-Dirac function and
$P^{a0}_{q}(k,k^{\prime})$ denotes the transition rate in
equilibrium.

Assuming that phonon-phonon scattering and phonon scattering by
impurities dominate over the phonon-carrier scattering
($\tau_{pc}\gg \tau_{ph}$), Eqs.(\ref{boltzmann}) and (\ref{N1nn})
give
\begin{equation}
\label{N1}
N^{1}_{q}=\frac{g_{s}g_{v}\tau_{ph}}{k_{B}T}\sum_{k,k^{\prime}}
\Gamma_{k^{\prime},k}\left(\frac{f^{1}_{k}}{df^{0}_{k}/dE_{k}}-
\frac{f^{1}_{k^{\prime}}}{df^{0}_{k^{\prime}}/dE_{k^{\prime}}}\right).
\end{equation}
In the above equation $f^{1}_{k}$ is the first order perturbation
of the carrier distribution function.

It is worth noting that Eq.~(\ref{N1}) can be regarded as a
starting point for the calculation of $S^{g}$ in all the problems
treated within the $\Pi$-approach\cite{Tsaousidoubook}. Now, by
substituting the phonon perturbation into (\ref{Q}) we take for
the heat flux
\begin{equation}
\label{heat}
Q=\frac{g_{s}g_{v}\tau_{ph}}{Lk_{B}T}\sum_{k,k^{\prime},q}\hbar\omega_{q}v_{q}
\Gamma_{k^{\prime},k}\left(\frac{f^{1}_{k}}{df^{0}_{k}/dE_{k}}-
\frac{f^{1}_{k^{\prime}}}{df^{0}_{k^{\prime}}/dE_{k^{\prime}}}\right).
\end{equation}

To determine the perturbation of the carrier distribution function
$f^{1}_{k}$ entering Eq.~(\ref{heat}) we use the 1D steady-state
Boltzmann equation
\begin{equation}
\label{boltzmann2} \frac{e}{\hbar}E\frac{\partial f_{k}}{\partial
k}=\left(\frac{\partial f_{k}}{\partial t}\right)_{coll},
\end{equation}
where $e$ is the carrier charge and the RHS of
Eq.~(\ref{boltzmann2}) is the rate of change of the carrier
distribution function due to elastic collisions with static
imperfections. In the relaxation time approximation this term is
written as $-f^{1}_{k}/\tau(E_{k})$ where $\tau(E_{k})$ is the
carrier relaxation time. Equation (\ref{boltzmann2}) is linearized
to give
\begin{equation}
\label{f1}
f^{1}_{k}=-eE\tau(E_{k})v_{k}\left(\frac{df^{0}_{k}}{dE_{k}}\right)
\end{equation}
where, $v_{k}=(1/\hbar)\nabla_{\bf k}E_{k}=\hbar k/m^{*}$ is the
carrier group velocity.

By substituting Eq.~(\ref{f1}) into (\ref{heat}) and making use of
Eqs.(\ref{M}), (\ref{onsager}) and (\ref{Gamma}) we finally get

\begin{eqnarray}
\label{SGS1} S^{g}=-\frac{g_{s}g_{v}e\tau_{ph}}{\sigma L
k_{B}T^{2}} \sum_{k,k^{\prime},q}\hbar\omega_{q}
v_{q}[\tau(E_{k})v_{k}-\tau(E_{k{^\prime}})v_{k^{\prime}}]\nonumber\\
\times f^{0}_{k}(1-f^{0}_{k^{\prime}})P^{a0}_{q}(k,k^{\prime}).
\end{eqnarray}
The above expression is equivalent to the expression derived by
Kubakaddi and Butcher\cite{Kubakaddi} for a quantum wire coupled
to 3D phonons. The authors in Ref.[\onlinecite{Kubakaddi}]
followed a different approach than this described here. They
followed Bailyn's theory~\cite{Bailyn} and they calculated the
phonon-drag contribution to the thermoelectric current that
originates from the carrier scattering with non-equilibrium
phonons in the presence of a small temperature gradient across the
wire. Their calculation was based on the solution of the coupled
equations for electrons and phonons.

The transition rate $P^{a0}_{q}(k,k^{\prime})$ is calculated by
using Fermi's golden rule. The lattice displacement for the
stretching mode is written in second quantized form
\begin{equation}
{\bf u(r)}=\sum_{q}\sqrt{\frac{\hbar}{2A\rho\omega_{q}}}\left
(\hat{{\bf \eta}}_{q}e^{iqz}\alpha_{q}+\hat{{\bf
\eta}}_{q}^{*}e^{-iqz}\alpha_{q}^{+}\right),
\end{equation}
where $\alpha_{q}^{+}$ and $\alpha_{q}$ are the phonon creation
and annihilation operators, respectively, $A=2\pi R L$ is the
nanotube surface area and $\rho$ is the mass density. For the
stationary carrier states considered here one easily finds
\begin{equation}
\label{rate}
P^{a0}_{q}(k,k^{\prime})=\frac{2\pi}{\hbar}N^{0}_{q}\frac{|U_{q}|^{2}}
{\epsilon^2(|q|,T)}\delta(E_{k^{\prime}}-E_{k}-\hbar\omega_{q})\,
\delta_{k^{\prime},k+q}
\end{equation}
where,  $N^{0}_{q}=[\exp(\beta\hbar\omega_{q})-1]^{-1}$ is the
phonon distribution in equilibrium, $|U_{q}|^{2}$ is the square of
the carrier-phonon matrix element for the deformation potential
coupling and $\epsilon(|q|,T)$ is the 1D static dielectric
function. By utilizing Eqs.~(\ref{def}) and (\ref{eta}) the matrix
elements $|U_{q}|^{2}$ in the limit $qR\ll 1$ are written as
\begin{equation}
|U_{q}|^{2}=\frac{\hbar\,\Xi^{2}q^{2}}{2A\rho\omega_{q}},
\end{equation}
where $\Xi=D(1-\nu)$. We note that the $q$-dependence of
$|U_{q}|^{2}$ is typical for the carrier interaction with
longitudinal acoustic phonons via an isotropic deformation
potential~\cite{Ridley}. A similar expression to the one we derive
here is given in Ref.~[\onlinecite{Pennington2}].

The dielectric function for a 1D gas confined to the surface of
the carbon nanotube is calculated within the random phase
approximation \cite{Fishman,Lin}. For the carrier wave functions
considered here we obtain
\begin{equation}
\label{dielectric}
\epsilon(|q|,T)=1+\frac{4g_{v}e^{2}m^{*}}{\hbar^{2}\pi
\epsilon_{b}}\frac{1}{|q|}K_{0}(|q|R)I_{0}(|q|R)M(|q|,T)
\end{equation}
where $I_{0}$ and $K_{0}$ are the modified Bessel functions of the
first and the second kind, respectively, and $\epsilon_b$ is the
background dielectric constant. $M(|q|,T)$ is the standard factor
that accounts for finite temperature effects on the static
polarization function~\cite{Fishman,Maldague}
\begin{equation}
\label{Factor} M(|q|,T)=\beta\int_{E_{1}}^{\infty}dE_{k}
\frac{\ln|(q+2k)/(q-2k)|}{4\cosh^{2}[\beta(E_{k}-E_{F})/2]}.
\end{equation}

To obtain an explicit expression for $S^{g}$ we substitute
Eq.~(\ref{rate}) into (\ref{SGS1}). Then the summation over
$k^{\prime}$ is readily carried out by replacing $k^{\prime}$ by
$k+q$ as a consequence of the momentum conservation condition
imposed by the Kronecker symbol $\delta_{k^{\prime},k+q}$.
Moreover, the summations over $q$ and $k$ are transformed to the
integrals
\begin{equation}
\sum_{q}\rightarrow \frac{L}{2\pi}\int_{-\infty}^{\infty}\,dq
\,\,\,\mbox{and}\,
\sum_{k}\rightarrow\frac{L}{2\pi}\int_{-\infty}^{\infty}
\,dk.\nonumber
\end{equation}
The presence of the $\delta$--function in Eq.~(\ref{rate}) allows
the immediate evaluation of the $k$--integration. We see by
inspection that
\begin{equation}
\label{delta}
\delta(E_{k+q}-E_{k}-\hbar\omega_{q})=\frac{2m^{*}}{\hbar^{2}|q|}\delta(2k+q\mp
q_{0})
\end{equation}
where, $q_{0}=2v_{s}m^{*}/\hbar$. The minus and the plus signs
correspond, respectively, to positive and negative $q$.

Now, after some algebra, we finally obtain
\begin{equation}
\label{SGF} S^{g}=\frac{m^{*}\Xi^{2}l_{ph}}{2\pi e \rho R
k_{B}T^{2}}\int_{0}^{\infty} dq\,
\frac{q}{\epsilon^{2}(|q|,T)}\frac{q}{2k_{F}}N^{0}_{q}\,I(q)
\end{equation}
where, $l_{ph}=v_{s}\tau_{ph}$ is the phonon-mean-free path,
$k_{F}=[2m^{*}(E_{F}-E_{1})/\hbar^{2}]^{1/2}$ is the Fermi wave
number and $I(q)$ is the product of the Fermi occupation factors
\begin{equation}
\label{IQ}
I(q)=f^{0}(E_{k})[1-f^{0}(E_{k}+\hbar\omega_{q})]
\end{equation}
with $k=|q_{0}-q|/2$. In deriving Eq.~(\ref{SGF}) we have ignored
the energy dependence of the carrier relaxation time and in
Eq.~(\ref{SGS1}) we have replaced $\tau(E_{k})$ by its value at
the Fermi level, $\tau_{F}$. This is a good approximation when
$\hbar\omega_{q}\ll E_{F}$ \cite{TBT}. Moreover, we have replaced
$\sigma$ by $ne^{2}\tau_{F}/m^{*}$ where $n=g_{s}g_{v}k_{F}/\pi$
is the density of carriers per unit length. Interestingly, $S^{g}$
becomes independent of the carrier relaxation time.

\subsection{An approximate expression for $S^{g}$} At low
$T$ and assuming that $\hbar\omega_{q}$ is a small quantity
compared to $E_{F}$ the product $I(q)$ is approximated by
\cite{CB}
\begin{equation}
\label{appr3}
I(q)\approx\hbar\omega_{q}(N^{0}_{q}+1)\delta(E_{k}-E_{F}),
\end{equation}
with $k=|q_{0}-q|/2$. The $\delta$-function can be written in the
following form
\begin{equation}
\delta(E_{k}-E_{F})=\frac{2m^{*}}{\hbar^{2}k_{F}}[\delta(q-q_{0}-2k_{F})
+\delta(q-q_{0}+2k_{F})].
\end{equation}
We see that $\delta(E_{k}-E_{F})$ resonates at $q=q_{0}+2k_{F}$
for positive $q$. When the expression (\ref{appr3}) for $I(q)$ is
substituted into Eq.~(\ref{SGF}) the integration over $q$ is
carried out straightforwardly by using the condition
$q=q_{0}+2k_{F}$. We note that $q_{0}\ll 2k_{F}$ and consequently,
stretching phonons with $q=2k_{F}$ make the dominant contribution
to $S^{g}$.

Equation (\ref{SGF}) is now significantly simplified and is
written in the convenient approximate form
\begin{equation}
\label{appr1}
S^{g}=\frac{C}{T^{2}}\frac{1}{\epsilon^{2}(2k_{F},T)}
\frac{e^{\beta\hbar\omega_{2k_{F}}}}
{(e^{\beta\hbar\omega_{2k_{F}}}-1)^{2}}
\end{equation}
where $C$ is given by
\begin{equation}
\label{C}
C=\frac{2(m^{*})^{2}\Xi^{2}l_{ph}\omega_{2k_{F}}}{\pi\hbar e\rho R
k_{B}}.
\end{equation}
In the above equations, $\omega_{2k_{F}}=v_{s}2k_{F}$ is the
frequency of a stretching phonon with $q=2k_{F}$ and
$\epsilon(2k_{F},T)$ is an approximate expression for the
dielectric function. To obtain $\epsilon(2k_{F},T)$ we replace $q$
by $2k_{F}$ in the denominator and in the arguments of the
modified Bessel functions $I_{0}$ and $K_{0}$ in
Eq.~(\ref{dielectric}). The factor $M(|q|,T)$ is replaced by the
average $\bar{M}(2k_{F},T)$ that is given by the expression
\begin{equation}
\label{aver} \bar{M}^{-2}(2k_{F},T)=\frac{\int_{0}^{\infty}
dq\,q^{2} M^{-2}(|q|,T)N^{0}_{q}I(q)}{\int_{0}^{\infty} dq\,q^{2}
N^{0}_{q}I(q)}.
\end{equation}
$\bar{M}(2k_{F},T)$ has been evaluated numerically for several
values of $k_{F}$ and $R$ and we find that in the degenerate limit
and when $T_{F}>5\hbar\omega_{2k_{F}}/k_{B}$ (where
$T_{F}=(E_{F}-E_{1})/k_{B}$ is the Fermi temperature) the
following expression provides a very good fit
\begin{equation}
\label{aver2}
\bar{M}(2k_{F},T)=\ln\left(\frac{4k_{F}+q_{0}}{q_{0}}\right)
[\alpha_{1}-\alpha_{2}\exp(-\alpha_{3}x)]
\end{equation}
where, $x=\beta\hbar\omega_{2k_{F}}$, $\alpha_{1}=1.175\pm 0.002$,
$\alpha_{2}=0.60\pm 0.01$ and $\alpha_{3}=0.41\pm 0.01$. At low
$T$ the effect of screening is severe and unity can be neglected
in Eq.(\ref{dielectric}). In this case the $T$-dependence of the
dielectric function is described by Eq.(\ref{aver2}).

At temperatures where $\beta\hbar\omega_{2k_{F}}\gg 1$ the
dielectric function shows a weak $T$-dependence. Then $S^{g}$
follows the law
\begin{equation}
\label{appr2}
S^{g}\propto\frac{1}{T^{2}}e^{-\beta\hbar\omega_{2k_{F}}}.
\end{equation}
This activated behavior is characteristic in 1D systems where the
Fermi surface consists of two discrete points $\pm k_{F}$
\cite{Campos,Lyo}.

We note that when screening is ignored and the phonon mean-free
path is constant the $T$-dependence of $S^{g}$ given by
Eq.~(\ref{appr1}) is similar to what predicted by Scarola and
Mahan \cite{Mahan} for an armchair (10,10) metallic SWCNT due to
interband electron scattering between the two linear bands.
However, the absolute magnitude of $S^{g}$ in a metallic tube is
expected to be much lower than that predicted in
Ref.[\onlinecite{Mahan}] due to the competing contributions of
electrons and holes to the thermoelectric current.

\section{Numerical results}
We assume that the free carriers are holes and we examine the
dependence of $S^{g}$ on temperature, the radius of the nanotube
and the position of the Fermi level with respect to the position
of the first van Hove singularity. The analysis is the same for
the case of electrons with the only difference being the sign of
$S^{g}$. The values for the material parameters used in the
calculations are $g_{s}=g_{v}=2$, $D=24$~eV~\cite{Stroscio,Ando},
$\nu=0.2$~\cite{Goldsman},
$\epsilon_{b}/4\pi\epsilon_{0}=2.4$~\cite{Lin}, $\rho=3.8\times
10^{-7}$~Kgr/m$^{2}$ and $v_{s}=19.9$~km/s~\cite{Ando}. The hole
effective mass is taken to be $m^{*}=m_{e}/22.7 \tilde{R}$ where
$\tilde{R}$ is the tube radius in nm~\cite{Goldsman}. We assume
that $l_{ph}=1$~$\mu m$.

\begin{figure}
\begin{centering}
\includegraphics[angle=0,height=9.0cm]{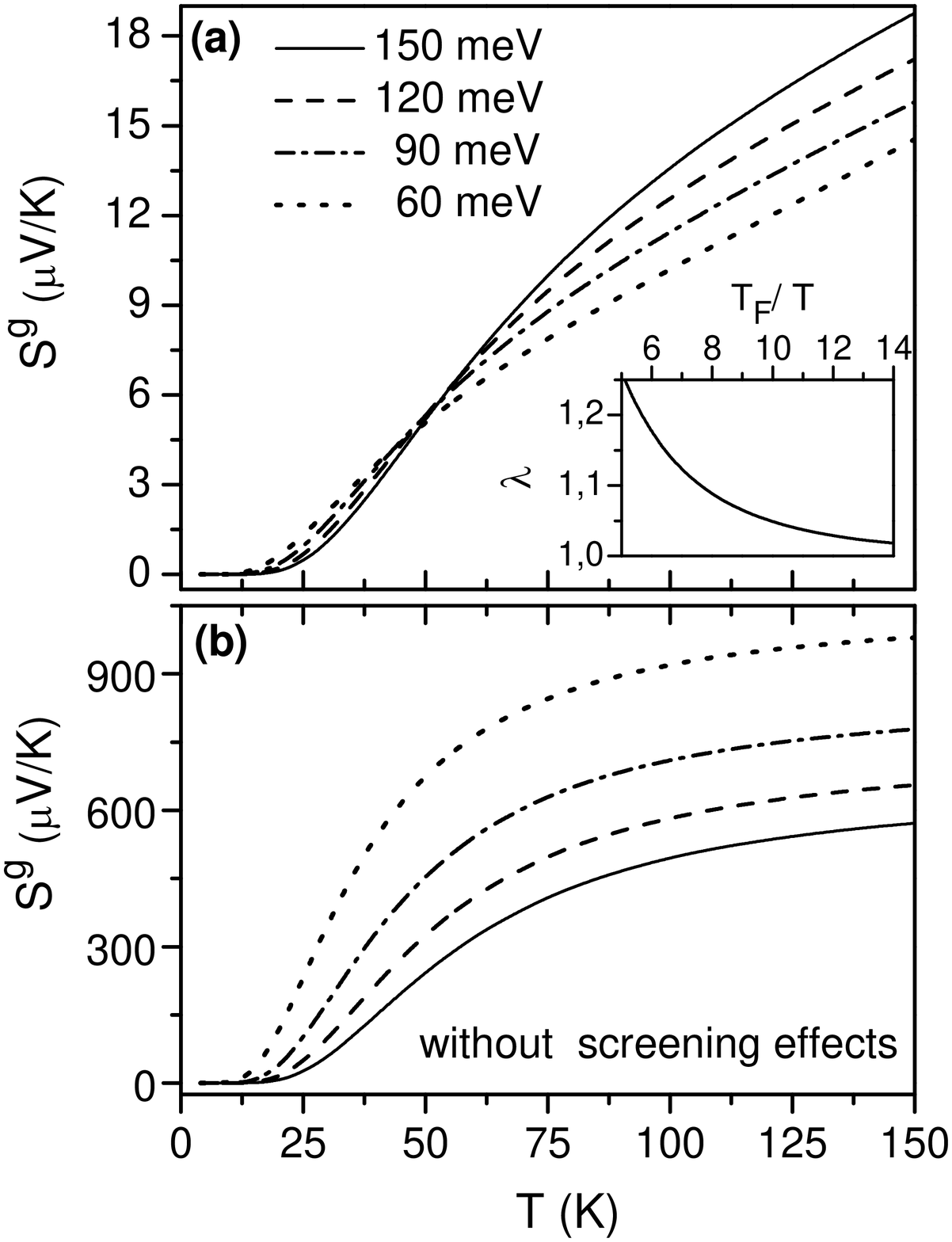}
\par\end{centering}
\caption{$S^{g}$ against temperature for a SWCNT of radius 0.5~nm.
Results are shown for four values of $E_{F}-E_{1}$: 60~meV(dotted
line), 90~meV (dashed-dotted line), 120~meV (dashed line) and
150~meV (solid line) when screening is taken into account (a) and
when screening is ignored (b). The phonon mean-free path is taken
to be 1~$\mu$m. The inset shows the ratio
$\lambda=S^{g}/S^{g}_{appr}$ as a function of $T_{F}/T$.}
\end{figure}

In Fig.~1a we show the $S^{g}$ evaluated from Eq.~(\ref{SGF}) for
a p-type SWCNT of radius $0.5$~nm as a function of $T$.  The
solid, dashed, dashed-dotted and dotted lines correspond,
respectively, to $E_{F}-E_{1}=150$, 120, 90 and 60~meV. We note
that at temperatures where the carriers are non-degenerate we have
taken into account the thermal broadening effects on $\sigma$. To
assure the accuracy of the approximate expression (\ref{appr1}),
in the inset of Fig.~1 we show the ratio
$\lambda=S^{g}/S^{g}_{appr}$ as a function of $T_{F}/T$ for
$E_{F}-E_{1}=60$~meV. $S^{g}$ and $S^{g}_{appr}$ are calculated
from Eqs.~(\ref{SGF}) and (\ref{appr1}), respectively.
Calculations of $\lambda$ for 90, 120 and 150~meV also fall on to
the same curve. We can see that in the degenerate limit the
approximate result agrees very well with the exact expression for
$S^{g}$. Finally, in Fig.~1b $S^{g}$ is calculated in the absence
of screening, $\epsilon(|q|,T)=1$. It turns out that screening
induces a strong suppression of $S^{g}$ by 1-2 orders of
magnitude. Inspection of Eq.~(\ref{dielectric}) shows that
screening effects become more severe as $R$ decreases. We note
that in the absence of screening $S^{g}$ levels off at high $T$ in
agreement with previous estimations in metallic
SWCNTs~\cite{Mahan,Fischer03}. However, when screening is
introduced $S^{g}$ shows a quasi-linear $T$-dependence at high $T$
due to the temperature dependence of the dielectric function. The
dielectric function $\epsilon(2k_{F},T)$ as a function of the
inverse temperature for a SWCNT with $R=0.5$~nm is shown in Fig.2

\begin{figure}
\begin{centering}
\includegraphics[angle=-90,width=9.5cm]{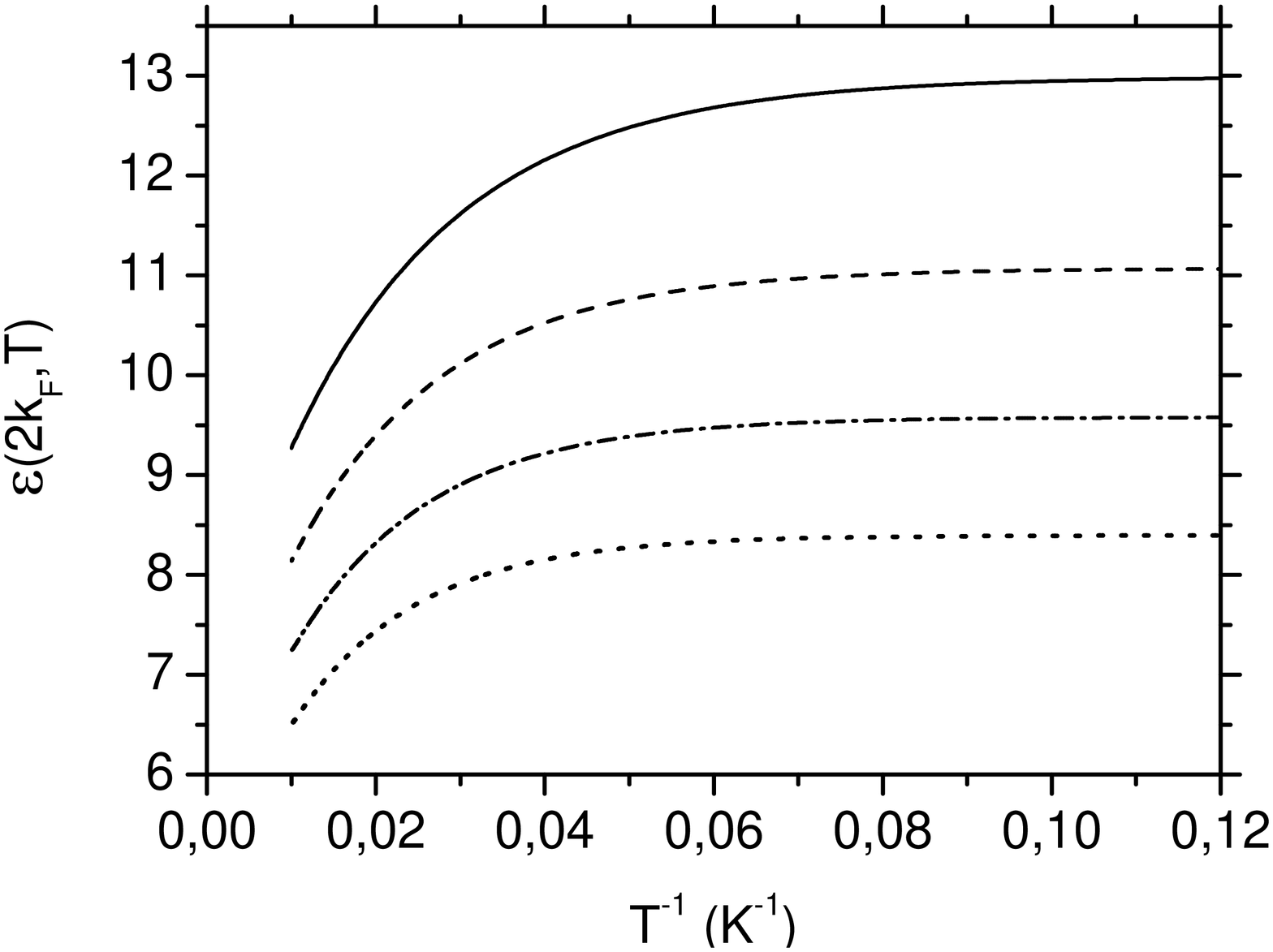}
\par\end{centering}
\caption{Dielectric function at $q=2k_{F}$ as a function of the
inverse temperature for a SWCNT of radius 0.5~nm. The solid,
dashed, dashed-dotted and dotted lines correspond to $k_{F}=0.4$,
0.45, 0.5 and 0.55 nm$^{-1}$, respectively.}
\end{figure}

In Fig.3 we show the dependence of $S^{g}$ on the Fermi level with
respect to the position of the first van Hove singularity for
temperatures $50\leq T\leq 300$~K. The shown structure is due to
two competing mechanisms which are the suppression of the
carrier-phonon scattering and the increase of $1/\epsilon(q)$ as
$k_{F}$ increases. The tube radius is 0.5~nm.

Finally, in Fig.4 we present the calculated values of $S^{g}$ as a
function of the nanotube radius. At temperatures higher than 100~K
we find that $S^{g}$ follows a law close to $S^{g}\propto
R^{-1.5}$. At lower temperatures $S^{g}$ shows a weaker dependence
on $R$ especially at large values of R.

\begin{figure}
\begin{centering}
\includegraphics[angle=-90,width=9.5cm]{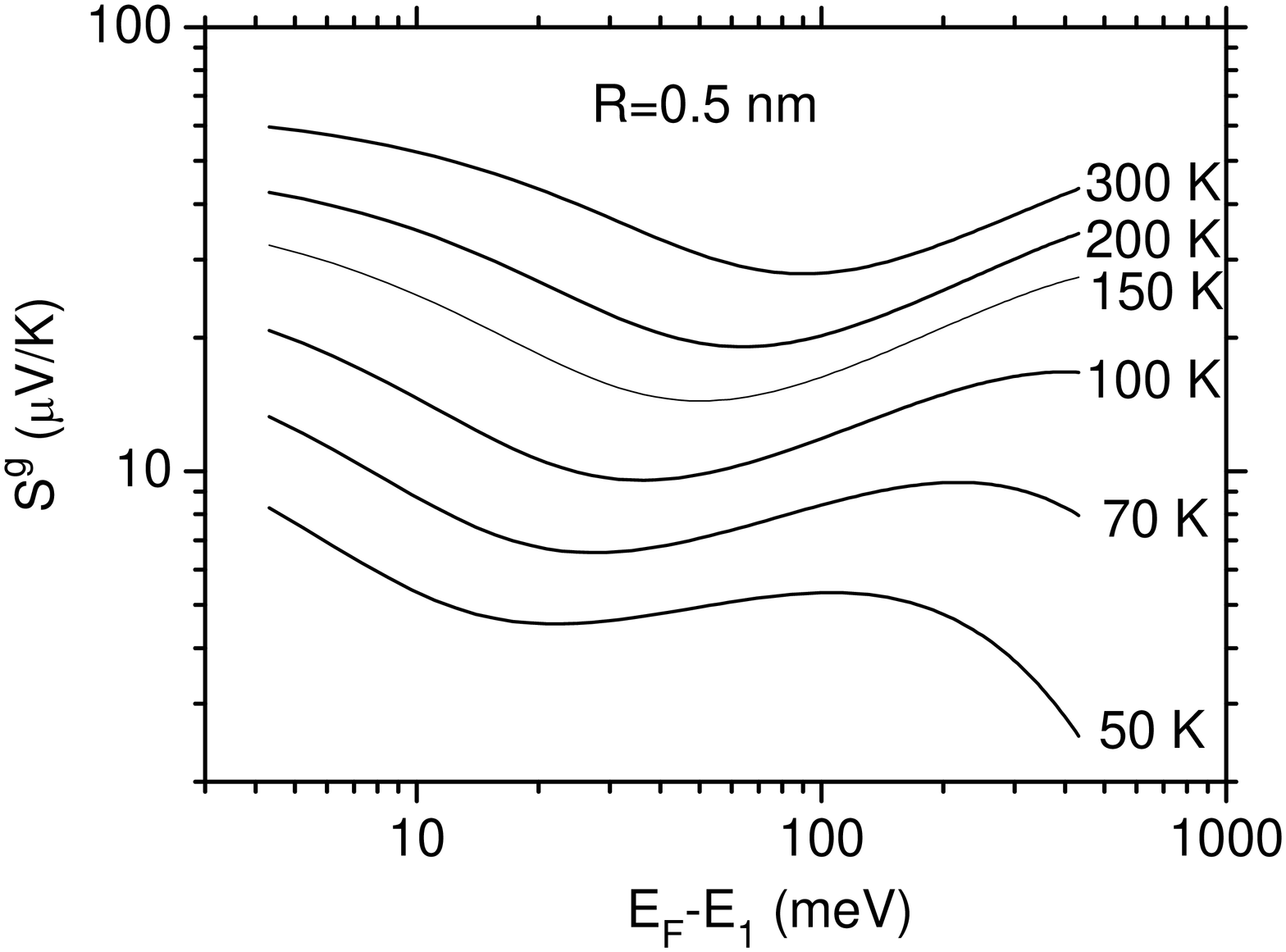}
\par\end{centering}
\caption{$S^{g}$ as a function of $E_{F}-E_{1}$ for various
temperatures. The phonon mean-free path is 1$\mu$m}
\end{figure}

\begin{figure}
\begin{centering}
\includegraphics[angle=-90,width=9.5cm]{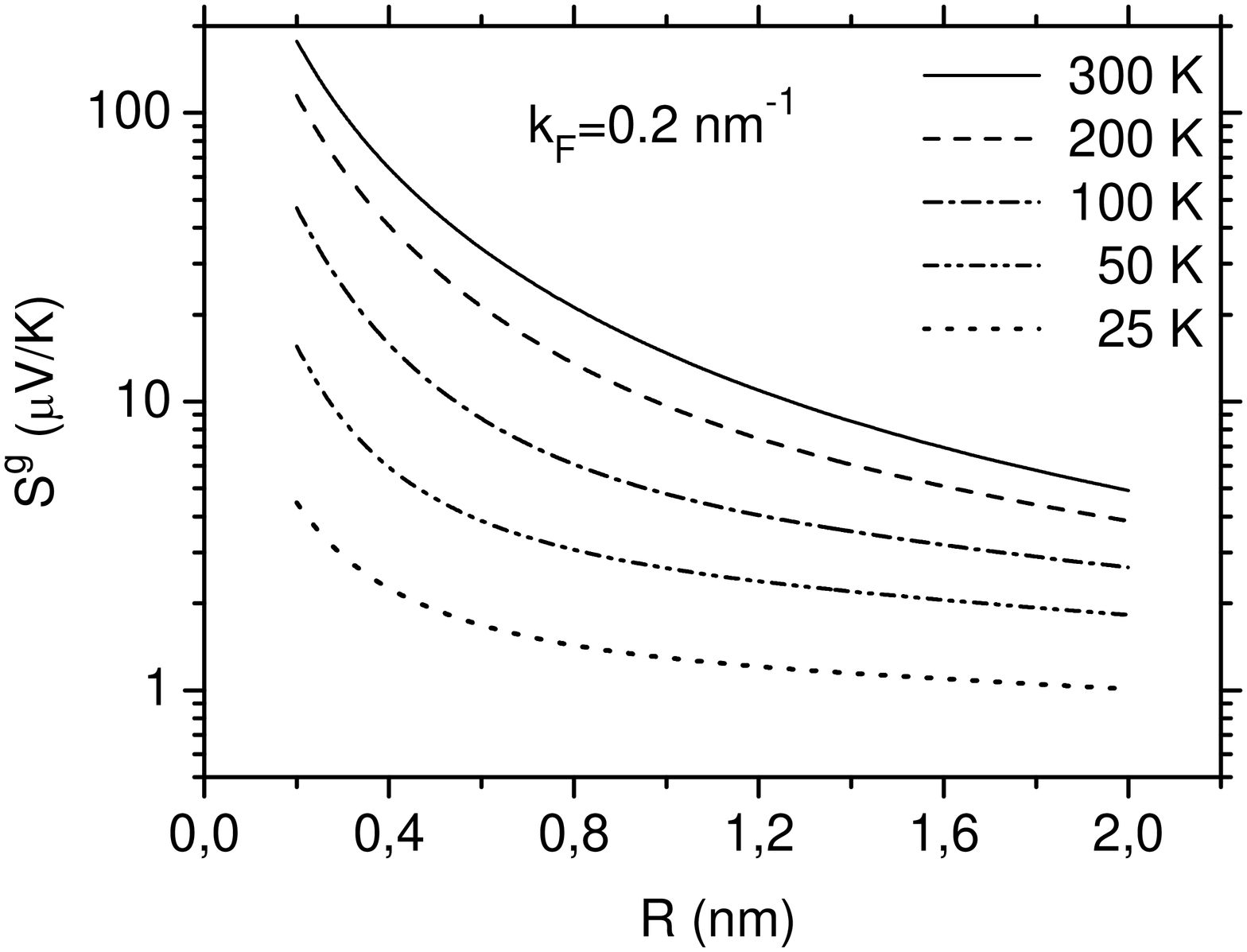}
\par\end{centering}
\caption{$S^{g}$ as a function of the nanotube radius for various
temperatures. $l_{ph}=1$~$\mu$m. At $T\geq 100$~K $S^{g}$ follows
approximately a $R^{-1.5}$ law.}
\end{figure}

\section{Comparison with the experiment-discussion}

So far there is no clear evidence about the phonon-drag effect in
isolated SWCNTs. The most relevant experiments were performed by
Yu et al.~\cite{Yu} in an individual SWCNT at temperatures above
100~K. The observed thermopower showed a linear $T$-dependence
which was attributed to the linear diffusion component and a
constant phonon-drag component of about 6~$\mu$V/K without,
however, excluding the possibility of an additional contact
effect. According to our analysis in Section III, the phonon-drag
thermopower at relatively high temperatures shows a quasi-linear
$T$-dependence and this makes difficult the separation of the
diffusion and the phonon-drag contributions. Nevertheless, when
the calculated values for $S^{g}$ shown in Fig.1a are fitted by a
linear function of $T$ we find that the intercepts vary from 1.4
to 7.4~$\mu$V/K when the position of the Fermi level with respect
to the first van Hove singularity varies from 60 to 150 meV. These
values are in agreement with the experimental estimation of
$S^{g}$ in Ref.[\onlinecite{Yu}]. We note that the intercepts
depend linearly on the phonon mean-free path and vary
approximately as $R^{-1.5}$.

Vavro {\em et al}.~\cite{Fischer03} and Zhou {\em et
al.}~\cite{Fischer05} have reported thermopower measurements in
p-doped bulk SWCNT samples in a wide temperature range (10-200~K)
that show clearly the signature of phonon drag. Normally, in bulk
samples nanotubes are self-organized into long \lq\lq ropes\rq\rq,
which contain a large number of nanotubes (tens to hundreds)
\cite{Thess}, forming a 3D network of complex geometry.
Thermopower in these nanotube networks exhibits a very similar
behavior as this of an individual nanotube described in Section
III. We have recently proposed a simple argument based on a model
of parallel conductors which suggests that in a network with
homogeneous doping and with a narrow distribution of tube
diameters the measured thermopower resembles that of an individual
tube \cite{Tsaousidoubook}. The resistivity measurements in the
samples under consideration showed weak coupling between metallic
nanotubes \cite{Fischer05} and hence the contribution from
metallic tubes to the total conductivity is neglected. We also
recall that the contribution of metallic tubes in $S$ is expected
to be small compared to this of semiconducting tubes. Therefore,
we can use the theory for isolate semiconducting SWCNTs developed
here to interpret the data in \cite{Fischer03,Fischer05}.

In Fig.5 the circles are the measured thermopower for a bulk
sample prepared by pulsed laser vaporization (PLV) and doped with
HNO$_{3}$\cite{Fischer05}. The tube radius is $R=0.68\pm 0.04$~nm.
At low temperatures (up to 100~K) we fit the data for the total
thermopower, $S$, by the expression
\begin{equation}
\label{fit}
S=\frac{C}{T^{2}}\frac{1}{\epsilon^{2}(2k_{F},T)}\frac{e^{\beta\hbar\omega_{2k_{F}}}}
{(e^{\beta\hbar\omega_{2k_{F}}}-1)^{2}}+AT(1-B\ln T).
\end{equation}
The first term is the approximate expression (\ref{appr1}) for
$S^{g}$ and the second term corresponds to the diffusion component
$S^{d}$. The sample is highly degenerate and at temperatures up to
100~K Eq.(\ref{appr1}) accurately describes $S^{g}$. The
$T$-dependence introduced by the dielectric function is given by
Eq.~(\ref{aver2}). The values we obtained for the parameters
$k_{F}$, $A$ and $B$ are shown in Table I.

The logarithmic term in $S^{d}$ secures an excellent fit to the
measured thermopower at all temperatures up to 100~K. If this term
is neglected the theoretical values for the total thermopower are
significantly larger than the experimental ones at high
temperatures. We speculate that the $T\ln T$ term in $S^{d}$ is
due to 2D weak localization (WL) effects \cite{Kearney}. If this
speculation is valid we would also expect a signature of WL in the
conductivity measurements. We note that the relative change in
conductivity should be the same as in $S^{d}$ but with an opposite
sign~\cite{Kearney,Lee}. Interestingly, we find that at
temperatures $10\leq T\leq 100$~K the conductivity follows the law
\begin{equation}
\label{sigma} \sigma=\sigma_{0}(1+B^{\prime}\ln T),
\end{equation}
where the value of $B^{\prime}$ is shown in Table I. We see that
$B$ and $B^{\prime}$ agree to each other remarkably well. The
origin of the 2D WL in these samples is not well understood. It is
likely related to the individual rope although in this case a 1D
localization behavior would be expected~\cite{Avouris}. However,
the phase coherence length, in the samples we discuss here, is
comparable to the diameter of the rope~\cite{FischerWL} and the 2D
limit might be approached. Langer {\em et al.}~\cite{Langer} have
also observed a $\ln T$ dependence of the conductance for an
individual multiwall CNT at 0.1-100~K which was attributed to 2D
WL. Finally, we should remark that WL is expected to have a
negligible effect on $S^{g}$ \cite{Miele}.

Now, by using the values for $k_{F}$ we obtained from the fitting
of the thermopower data at low $T$ we calculate $S^{g}$ in the
whole temperature range from 10 to 200~K by using the exact
expression (\ref{SGF}). The only remaining unknown is the value of
the phonon-mean-free path $\l_{ph}$ which is determined from the
experimental data when the diffusion contribution is subtracted.
We find that $l_{ph}=0.6$~nm. This value is consistent with the
values 0.25-0.75~$\mu$m reported recently for an individual
SWCNT~\cite{Yu}. Our estimation for the total thermopower is shown
as solid line in Fig.5. The dashed and the dashed-dotted lines
correspond to the phonon-drag and the diffusion contributions,
respectively.

By following a similar procedure as this described above we have
interpreted the thermopower data for another bulk sample prepared
by high pressure decomposition of CO (HiPco) and doped with
H$_{2}$SO$_{4}$ \cite{Fischer03}. The tube radius varied from
0.4-0.7 nm. Conductivity measurements for this sample (designated
as HPR93C) appear in \cite{Zhou3}. The experimental data for the
ratio $S/T$ are shown as squares in Fig.6. The values for the
fitting parameters $k_{F}$, $A$ and $B$ are shown in Table I. We
also present the value of $B^{\prime}$ for comparison. In the
calculations the tube radius is taken to be the average
$R=0.55$~nm while for the phonon-mean-free path we obtained the
value $l_{ph}=0.4$ nm. The calculated values for $S/T$ is shown as
solid line in Fig.6.

\begin{figure}
\begin{centering}
\includegraphics[angle=-90,width=10.5cm]{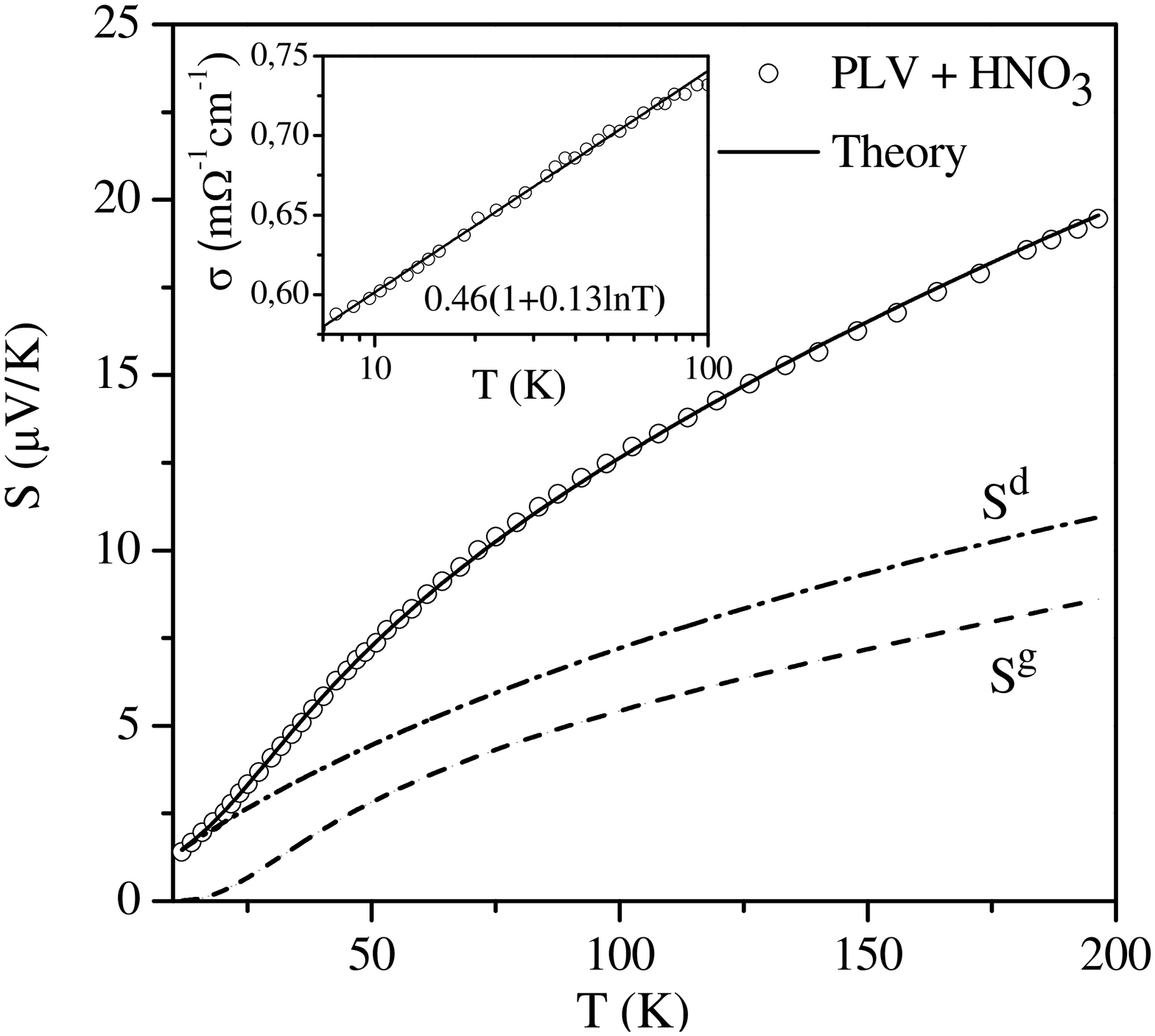}
\par\end{centering}
\caption{Thermopower versus temperature. The circles are the
experimental data for a bulk sample prepared by pulsed laser
vaporization and doped with HNO$_{3}$\cite{Fischer05}. The solid
line is the total thermopower of an individual nanotube obtained
as explained in text. The dashed and the dashed-dotted lines
correspond to the phonon-drag and the diffusion contributions,
respectively. In the inset the circles are the conductivity data
\cite{Fischer05} and the solid line is the fit by using
Eq.~(\ref{sigma}).}
\end{figure}

\begin{table}[ht]
\caption{The values for the parameters $A$, $B$ and $k_{F}$
obtained from the fit of the thermopower data
\cite{Fischer03,Fischer05} for $T\leq 100$~K by using
Eq.~(\ref{fit}). In the last column we show for comparison the
values for $B^{\prime}$ obtained from the resistivity data
\cite{Fischer05,Zhou3} in the range 10-100~K.}
\begin{ruledtabular}
\begin{tabular}{cccccc}
\hline
 &$A$\,($\mu V/K^{2}$)&$k_{F}$\,($nm^{-1})$&$B$&$B^{\prime}$\\
\hline
PLV film+HNO$_{3}$& 0.184$\pm$0.003 & 0.40$\pm$0.01& 0.132$\pm$0.002&0.131$\pm$0.002\\
HiPco fiber+H$_{2}$SO$_{4}$& 0.083$\pm$0.007 & 0.57$\pm$0.01&0.156$\pm$0.013&0.222$\pm$0.003\\
\hline
\end{tabular}
\end{ruledtabular}
\end{table}

Concerning the consistency of the fitting parameters $A$ and
$k_{F}$ we should make the following remarks. By using the values
for $k_{F}$ shown in Table I and a simple tight binding model for
the estimation of the first van Hove singularity (see, for
example, Ref.~[\onlinecite{Goldsman}]) the values we get for
$E_{F}$ are in good agreement with those determined from
reflectivity and Raman measurements~\cite{Fischer05}. Also, $A$
varies inversely with $k_{F}^{2}$ in agreement with Mott's
expression for $S^{d}$. Moreover, the values we extract for
$k_{F}$ support recent arguments according to which
H$_{2}$SO$_{4}$ is a stronger dopant than HNO$_{3}$
\cite{Fischer05}. Namely, according to our estimation for the
Fermi wave numbers, the Fermi level is shifted by 94 and 155~meV
below the top of the valence band for the PLV+HNO$_{3}$ and
HiPco+H$_{2}$SO$_{4}$ samples, respectively.

\begin{figure}
\begin{centering}
\includegraphics[angle=-90,width=9.5cm]{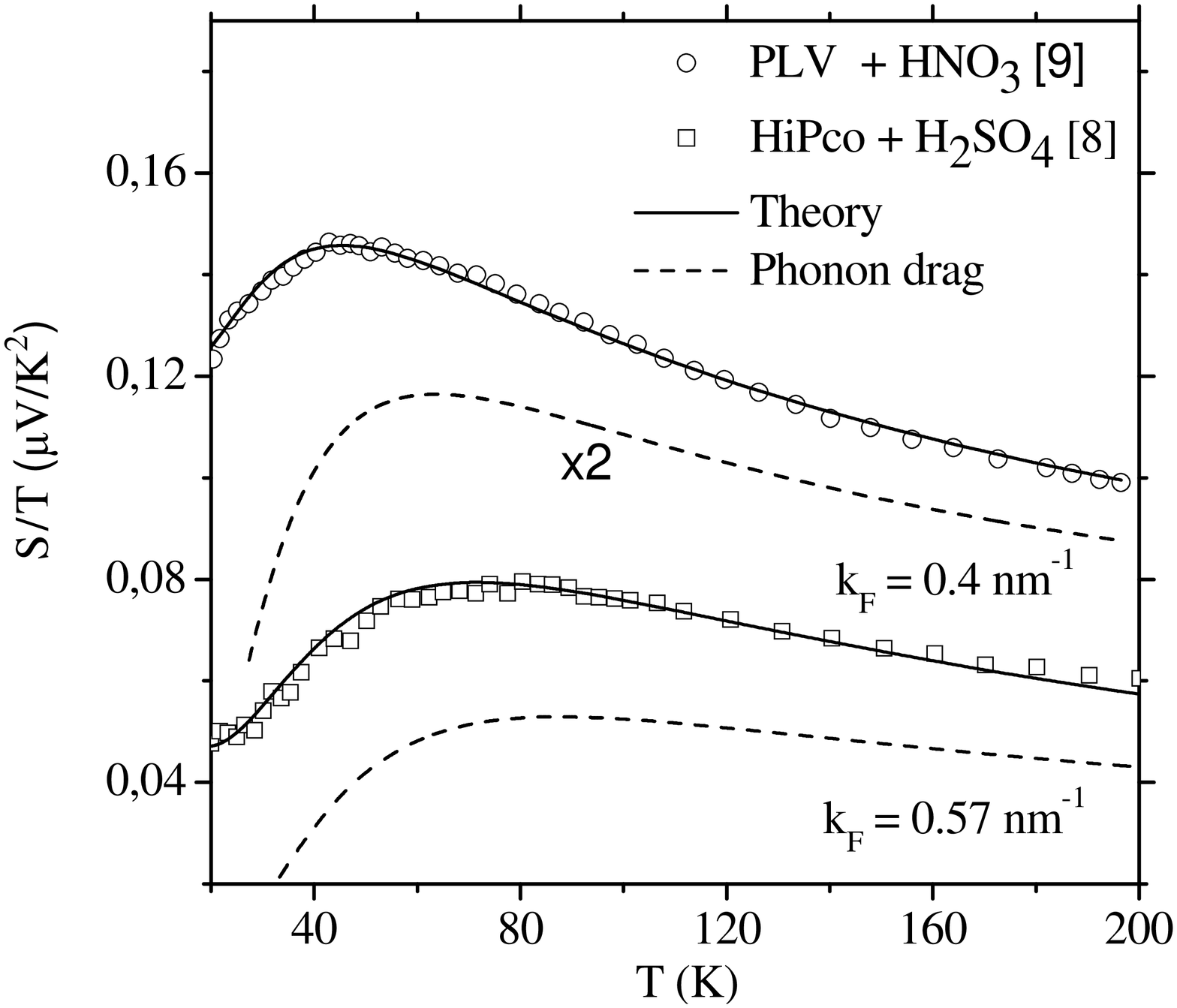}
\par\end{centering}
\caption{The ratio S/T as a function of temperature. The symbols
are the experimental data for two bulks samples
\cite{Fischer03,Fischer05}. The solid lines denote the total
thermopower of an individual nanotube obtained as explained in
text. The dashed lines are the phonon-drag contributions. For
clarity the calculated $S^g$ that corresponds to the PLV+HNO$_{3}$
sample has been multiplied by the factor 2. The peaks shown in the
measured $S/T$ are associated to the phonon-drag effect.}
\end{figure}

In order to show clearly the effect of phonon drag in Fig.6 we
have plotted the ratio $S/T$ as a function of $T$. The circles and
the squares are the measured values for the samples PLV+HNO$_{3}$
and HiPco+H$_{2}$SO$_{4}$, respectively. The dashed lines are the
theoretical estimates for $S^{g}$ and the solid lines are the
calculated values for the total thermopower. The peaks at
$T=T^{*}$ are associated to phonon-drag thermopower. The shift
between the experimental and the theoretical values for $T^{*}$ is
due to the logarithmic term in $S^{d}$. The position of the peak
moves towards to higher temperatures as doping increases. This
dependence can be understood by maximizing the ratio $S^{g}/T$
using Eq.~(\ref{appr1}). Then we get the following dependence
\begin{equation}
\label{Tstar}
T^{*}=1.1\frac{\hbar v_{s}k_{F}}{k_{B}}.
\end{equation}

It is important to add that the exponential suppression of $S^{g}$
at low temperatures is unique for 1D systems. In higher dimensions
$S^{g}$ exhibits a power-law $T$ dependence at low
temperatures~\cite{Fletcher04,Tsaousidoubook}. The observed peak
in $S/T$, which is ascribed to phonon drag, underlies the 1D
character of thermopower. This adds another confirmation that
thermopower in bulk carbon nanotube-based materials is a property
of the individual tube rather than a property of the network.

\section{Conclusions}
In summary, we have presented a rigorous model for the calculation
of the phonon-drag thermopower in degenerately doped
semiconducting SWCNTs. By using the derived model we investigated
the dependence of $S^{g}$ on temperature, tube radius and position
of the Fermi level. We found that $S^{g}$ decreases with the
increase of the tube radius following approximately a $R^{-1.5}$
law at high temperatures. In the degenerate limit, we derive a
simple expression for $S^{g}$ which can be used as a probe for the
estimation of the free carrier density in doped tubes. According
to this expression $S^{g}$ shows an activated $T$ dependence at
low temperatures. Screening effects of the carrier-phonon coupling
reduce the magnitude of $S^{g}$ severely and result to a
quasi-linear $T$-dependence of phonon drag at high $T$. Finally,
we have compared our model with available data in acid-doped bulk
samples~\cite{Fischer03,Fischer05} and we found a very good
agreement in a wide temperature range.

\section*{Acknowledgements}
The author wishes to thank Dr. K. Papagelis for extensive and
useful discussions and Prof. R. Fletcher for stimulating remarks.

\end{document}